\documentclass[a4paper,12pt]{article}

\usepackage[a4paper]{geometry}
\usepackage[dvipsnames]{xcolor}
\usepackage[utf8]{inputenc} 
\usepackage{hyperref}
\usepackage{float}
\usepackage{amssymb,amsmath,amsthm}
\usepackage{amsfonts}

\pdfoutput=1
\usepackage{graphicx}
\usepackage{graphicx, float, tabularx, booktabs}
\usepackage{color}
\usepackage{enumerate}
\usepackage[american]{babel}
\usepackage{stackrel}

\usepackage[numbers, compress]{natbib}
\usepackage{natbib}
\newcommand{\f}{\frac}

\newcommand{\bq}{\begin{eqnarray}}
\newcommand{\fq}{\end{eqnarray}}
\newcommand{\be}{\begin{equation}}
\newcommand{\ee}{\end{equation}}
\newcommand{\bea}{\begin{eqnarray}}
\newcommand{\eea}{\end{eqnarray}}
\newcommand{\mpl}{M_{\rm P}}

\newcommand{\lpl}{L_{\rm P}}

\usepackage{amsfonts}
\usepackage{epsfig}

\usepackage{graphicx, tabularx}
\newcommand{\beq}{\begin{equation}}
\newcommand{\eeq}{\end{equation}}
\newcommand{\diff}{\mathrm{d}}
\renewcommand{\d}{\diff}

\vspace{-0.1cm}
\renewcommand{\thefootnote}{\fnsymbol{footnote}}

\author{Bernard Carr$^{a}$\thanks{b.j.carr@qmul.ac.uk}, \
Piero Nicolini$^{b,c}$\thanks{nicolini@fias.uni-frankfurt.de} \ and
Athanasios G. Tzikas\thanks{tzikas@fias.uni-frankfurt.de}
 \\[1ex]
\small $^a$ School of Physics and Astronomy, Queen Mary University of London, 
 \\[-0.5ex]
\small  Mile End Road, London E1 4NS, UK \\[1ex]
\small $^b$ Frankfurt Institute for Advanced Studies (FIAS)\\[-0.5ex]
\small Ruth-Moufang-Str.~1, D-60438 Frankfurt am Main, Germany
\\[1ex]
\small $^c$ Institut f\"{u}r Theoretische Physik, Johann Wolfgang Goethe-Universit\"{a}t Frankfurt\\[-0.5ex]
\small Max-von-Laue-Str.~1, D-60438 Frankfurt am Main, Germany\\
[1ex]
}
\date{{}}
\title{Quantum gravity  black holes  as dark matter?}
\begin{document}

\begin{center}
{\Large\bf  
Quantum gravity  black holes  as dark matter?}

\end{center}

\vspace{-0.1cm}

\begin{center}

Bernard Carr
$^{1}$\footnote{b.j.carr@qmul.ac.uk},
Piero Nicolini
$^{2,3,4}$\footnote{piero.nicolini@units.it},
Athanasios G. Tzikas
$^{3,5}$\footnote{athanasios.tzikas@unibg.it}

\vspace{.6truecm}

{ $^1$School of Physics and Astronomy, Queen Mary University of London,\\
Mile End Road, London, E1 4NS, UK. }\\ \vspace{1.5mm}

{ $^2$Dipartimento di Fisica, Università degli Studi di Trieste and Istituto \\ Nazionale di Fisica Nucleare (INFN), Sezione di Trieste,
\\
Strada Costiera 11,  Trieste, 34151, Italy.}\\ \vspace{1.5mm}

{ $^3$Institut f\"ur Theoretische Physik, Johann Wolfgang Goethe-Universität \\ Frankfurt am Main and Frankfurt Institute for Advanced Studies, \\
Ruth-Moufang-Str. 1,  Frankfurt am Main, 60438, Germany.}\\ \vspace{1.5mm}

{ $^4$Center for Astro, Particle and Planetary Physics, New York University\\ Abu Dhabi, P.O. Box 129188, Abu Dhabi, United Arab Emirates.}\\ \vspace{1.5mm}

{ $^5$Dipartimento di Ingegneria e Scienze Applicate, Università degli Studi\\
 di Bergamo,Viale Marconi 5,  Dalmine, 24044, Italy.}\\

\end{center}

\vspace{1mm}

\begin{abstract}
\noindent 
{One of the major problems in quantum gravity research is the lack of signals at the reach of present or near-future experimental facilities. In this paper, we show that this is not the case. Contrary to previous claims, the quantum decay of de Sitter space into black hole spacetimes can be significant even after inflation and can be observed  on galactic scales.
Using the instanton formalism within the no-boundary proposal for a class of short-scale, quantum-gravity-improved black holes, we show that de Sitter space decay would result in the production of $10^{60}$ stable Planck-size black hole remnants  within the current Hubble  horizon, which is the number required to explain dark matter.}
\noindent
   
\end{abstract}

\renewcommand{\thefootnote}{\arabic{footnote}}
\setcounter{footnote}{0}

\newpage
\section{Introduction}\label{sec:intro}

Since Hawking's pioneering work on the thermal radiation of black holes, research in gravitational physics has made enormous progress. However, the field of quantum gravity remains problematic, mainly due to the coexistence of competing theoretical formulations and the lack of experimental data \cite{Nicolai13}.
The severity of this situation has been minimized by believing that quantum gravity has a narrow range of applications, including only the early stages of the universe and the physics of microscopic black holes, with no observational consequences. 
In this paper we 
show that this is not the case by demonstrating that there is a quantum gravity phenomenon that has a global, long-range effect on the structure of the universe.
This phenomenon is the quantum mechanical decay of de Sitter space.

Within the so called ``no-boundary proposal'',  any universe can be described by a wavefunction $\Psi$ \cite{HaH83}, 
given by the path integral over all possible Euclidean metric configurations that can be analytically continued to the 
Lorentzian section \cite{GiH77}. 
This can be evaluated with the saddle point method,
\begin{equation} \label{eq:wvf}
\Psi \approx e^{-I} ,
\end{equation}
where 
$I$ is 
the instanton action.
In their seminal works, Mann and Ross \cite{MaR95} and  also Bousso and Hawking \cite{BoH95,BoH96} evaluated  this
for Reissner-Nordström-de Sitter spacetime and Schwarzschild-de Sitter spacetime, respectively. Their results showed that the decay rate of de Sitter space into a universe containing a black hole is either negligible or has no observational consequences after inflation.

 Our first point 
 is that this negative conclusion is
due to an underestimate of the gravitational amplitude. This is because classical spacetimes cannot generate all possible instanton configurations. 
In the Schwarzschild-de Sitter case, for example, only the degenerate Nariai black hole contributes to the action. When there is electric charge, the spacetime has an extra horizon that allows for more instantons, 
namely lukewarm, cold, and ultracold instantons. However, their contribution is suppressed by a surface term that contains the gauge field \cite{MaR95,HaR95}.
Short-scale quantum gravity corrections to black hole spacetimes also result in an additional horizon, regardless of the specific quantum gravity formulation (see, for instance, \cite{BoR00, NSS06, NSW19}) or whether the classical singularity is cured or only softened \cite{IMN13}.
Indeed,
quantum-gravity corrections 
change the horizon topology even for the neutral, static case \cite{MaN11}. 
Furthermore,  the gauge field vanishes there, which ensures that there is no suppression by surface terms.

 Secondly,  although the post-inflationary production rates of quantum-gravity improved spacetimes are small, they are not infinitesimal and  this  can explain the dark matter content in terms of Planckian black hole remnants. 
We emphasize that this
 is possible only because of the horizon topology change, which solves both the production and stability problems 
of the classical black hole spacetimes. 
Indeed, quantum-gravity-improved black holes do not undergo significant decay due to Hawking or Schwinger mechanisms.  Rather, in the final stages of evaporation, they cool down to equilibrium with the environment, which  has
close to zero temperature. 



In summary, our proposal provides a physical mechanism for producing black holes and explaining dark matter, independent of any particular model of short-scale gravity modifications (see, for example, \cite{Dym92,BMMN19,ACS01,CMN15,CPV25,DEL25}).
Our message is that any modification of classical black hole metrics at very short length scales causes a change in instanton topology, resulting in phenomenological consequences for the universe on galactic scales.

\section{Results}\label{sec:results}

The decay rate can be obtained by squaring \eqref{eq:wvf} and calculating the ratio of the probability of the universe with an object to that of the empty background:
\begin{equation} \label{eq:rate}
\Gamma = \f{|\Psi_{\rm obj}|^2}{|\Psi_{\rm bg}|^2} = \exp \left[  -2 (I_{\rm{obj}}-I_{\rm bg}) \right]  .
\end{equation}
Since the Gibbons-Hawking-York term does not contribute, because the boundary has vanishing extrinsic curvature in the Euclidean section \cite{MaR95,MaN11}, one can calculate the instanton action from the Wick-rotated form of the gravitational-matter action in Planck units:
\begin{equation} \label{eq:action}
I=- \int \d ^4 x \sqrt{g} \left( \f{R-2\Lambda}{16\pi} + \mathcal{L}_{\rm m} \right) .
\end{equation}
Setting the matter Lagrangian $\mathcal{L}_{\rm m}$  to zero in \eqref{eq:action}, and expressing the Ricci scalar $R$ in terms of the cosmological constant $\Lambda$, yields 
\begin{equation} \label{eq:ds_I}
I_{\rm bg} = - \f{3\pi}{2\Lambda} \,
\end{equation}
for the empty de Sitter background.   
 However, $\mathcal{L}_{\rm m}\neq 0$ for spacetimes containing objects such as black holes \cite{MaN11}.
In the static, neutral case
 the 
 metric is
\begin{equation} \label{eq:line_el}
\d s^2 = -f(r) \d t^2 + f^{-1}(r) \d r^2 + r^2 \d \Omega ^2 ,
\end{equation}
with $\d \Omega^2 = \d \theta ^2 + \sin ^2 \theta \ \d \phi ^2$ and
\begin{equation} \label{eq:potential}
f(r)=1-\f{2m(r)}{r} - \f{\Lambda}{3} r^2 .
\end{equation}
The function $m(r)$ is 
the cumulative mass distribution. 
For classical black holes, 
 it is just a constant $M$.  However, when quantum effects are included, $m(r)$ takes the general form of the integral of the mixed time-time component of the energy-momentum tensor. 
This tensor is non-vanishing because quantum effects  generally prevent
 complete gravitational collapse and spread the mass density over some distance $\ell$. In order to satisfy 
$m(r)\to M$ at infinity, it is convenient to shift the cosmological term into the Einstein tensor.
The energy-momentum tensor then has the universal form
\begin{equation} \label{eq:tensor}
T^{\mu}_{\nu} = \mathrm{diag}\left( -\rho(r),p_r(r),p_{\bot}(r),p_{\bot}(r) \right)  ,
\end{equation}
whatever theory is behind the short scale modification of general relativity. If we take the limit $\ell\to 0$, the energy-momentum tensor only exists in a distributional sense, and one recovers the classical Schwarzschild-de Sitter solution  with  $m(r)=M$ \cite{BaN93}. 

Quantum gravity enters the picture
by imposing a certain profile for the  components
of the energy-momentum tensor. The scale $\ell$ is generally free, like  the parameter $\alpha^\prime$ in string theory, but it makes sense to set it close to the Planck length $\lpl $.
A very interesting property is that, regardless of the specific profile of $\rho(r)$, $p_r(r)$ and $p_{\bot}(r)$, for values of $M$ in the interval between a threshold $M_\mathrm{c}(\Lambda)$ and a mass $M_\mathrm{N}(\Lambda)$, 
the geometry allows a Cauchy horizon $r_1$ in addition to the event  horizon $r_2$ and cosmological horizon $r_3$. 
At the endpoints of the interval, 
$M=M_\mathrm{c}$ and $M=M_\mathrm{N}$, two of the 
three horizons merge, corresponding to the cold (c) and Nariai (N) degenerate configurations.
However, for  masses
outside this
interval, i.e. for $M<M_\mathrm{c}$ and for $M>M_\mathrm{N}$, there is only a cosmological horizon and no black holes form.
 This contrasts with general relativity, where black holes can exist for any $M$.
\begin{figure}[!t] %
\centering
\includegraphics[height=2.2in]{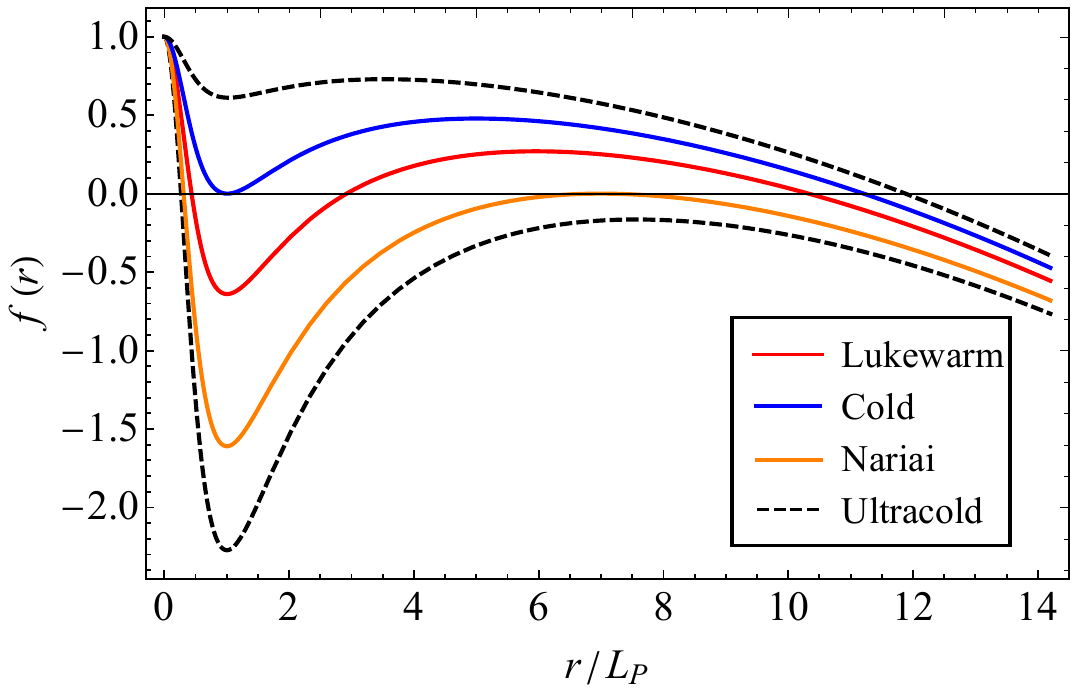}%
\caption{The metric potential of the Euclidean manifolds (gravitational instantons) of a regular black hole in de Sitter space \cite{NSS06, MaN11} for $\ell = \lpl$. Increasing the black hole total mass $M$ corresponds to moving 
along the curves for constant $\Lambda$. Single-horizon configurations (dashed curves) 
correspond to an ultracold instanton.}
\label{fig:fig1}
\end{figure}
One finds four possible instantons:
ultracold, Nariai, cold and lukewarm (lw), corresponding to  single-horizon,
two degenerate double-horizon
and non-degenerate triple-horizon configurations, respectively $-$ see Fig.~\ref{fig:fig1}.

We now have all the necessary elements to calculate the decay rate $\Gamma$ in \eqref{eq:rate}. 
 We just focus on the lukewarm case because the rates of the other instantons
are negligible compared to this. 
 In the lukewarm case, 
we satisfy the 
thermal equilibrium condition by  requiring
equal surface gravities $\kappa_2=\kappa_3$ at the
$r_2$ and $r_3$.   This implies
\begin{equation} \label{eq:lw_I}
I_{\rm lw} = - \f{\Lambda \beta}{4} \left( r_3^3 - r_2^3 \right) - \f{\beta}{2} \big[r_3\ m'(r_3)-r_2\ m'(r_2)-r_3+r_2 \big] \,, 
\end{equation}
where $\beta$ 
indicates the instanton periodicity,   i.e.
the inverse of the Hawking temperature, and $m^\prime(r_i)$ is the radial derivative of $m(r_i)$ for $i=2, 3$.
From the above definition of cumulative mass,  this derivative is simply $4\pi \rho(r_i)$.

If we assume some
value for the total mass, $M=M_\textrm{lw}$ with $M_\mathrm{c}<M_\textrm{lw}<M_\mathrm{N}$,  then due to the dependence of 
$M_\mathrm{c}$ and $M_\mathrm{N}$ on $\Lambda\,$, the lukewarm  instanton exists only if the cosmological constant  is
less than its
value in the Nariai configuration, i.e. $\Lambda < \Lambda_\mathrm{N}$.
From \eqref{eq:lw_I} it is clear that 
the two horizons are of comparable size, $r_2\simeq r_3\,$, for values  just below $\Lambda_\mathrm{N}$.
So the instanton  action tends to zero  and
the resulting rate is exponentially suppressed, $\Gamma_{\rm lw} \sim e^{-3\pi/\Lambda}$.

Due to the causal structure of the instanton, 
the 
radii $r_2$ and $r_3$ can be very different for smaller values of $\Lambda$.  In the limit $\Lambda \ll \Lambda_\mathrm{N} \sim 1$ (e.g. after inflation), the black hole shrinks to the  Planck
scale from above  while $r_3$ tends to de Sitter radius $r_\mathrm{dS}=\sqrt{3/\Lambda}$ from below.  As a result, for $r_2\ll r_3\,$, the rate \eqref{eq:rate} for the lukewarm  instanton \eqref{eq:lw_I} is
\begin{equation} \label{eq:lwrate2}
\Gamma_{\rm lw} \approx \exp\left[ 2\pi  \left( \f{3(r_3^2/r^2_\mathrm{dS})-2}{3(r_3^2/r^2_\mathrm{dS})-1} \right)r_3^2 - \pi r_{\rm dS}^2  \right]. 
\end{equation}
The case $r_3=r_\mathrm{dS}$ is forbidden because it corresponds to $M_\textrm{lw}=0\,$, so the lukewarm rate \eqref{eq:lwrate2} tends to unity from below $\Gamma_{\rm lw}\lesssim 1$. 

\section{Discussion}\label{sec:discussion}

The first remark about 
 \eqref{eq:lwrate2} is that, contrary to previous claims \cite{MaR95,BoH95,MaN11},  de Sitter decay can be non-negligible even for small, sub-Planckian values of the cosmological constant ($\Lambda\ll 1)$.
This result follows 
from the presence of $m(r)$ instead of the total mass $M$ in \eqref{eq:potential} and is universal, i.e., it determines the horizon topology change irrespective of the specific theory behind the cumulative mass distribution.
Since the normalized decay probability $\mathcal{P}\equiv \Gamma/(1+\Gamma)$  can be close to but less than 50\%, the energy associated with the cosmological constant cannot be completely converted into black holes.
From this perspective, the universe can be considered quantum mechanically stable 
today.
 
In order to calculate the actual number of black holes from the early epochs of the universe, we must properly define the spacetime to which the rate \eqref{eq:rate} refers. We must also determine the number density of black holes per time unit, which differs  dimensionally from the 
parameter $\Gamma$.
Additionally, we must take into account the cosmological expansion.
To address  these
 issues, we 
refer to the work of Bousso and Hawking \cite{BoH96}, which is based on the following hypotheses:
\begin{enumerate}
\item A time-dependent cosmological constant.
\item The existence of Hubble-size domains, or ``bubbles,'' that expand independently from their surroundings.   
\item The occurrence of topological fluctuations that allow for black hole production in each Hubble-size domain. 
\end{enumerate}
Within this framework, Bousso and Hawking treat each bubble as a miniature de Sitter universe that decays into a black hole spacetime at a rate $\Gamma$. This determines
 the number of black holes produced per 
Hubble time in a Hubble volume.
It also allows us to determine the total number of black holes during the expansion of the universe by counting the bubbles within the Hubble horizon.
The formalism is derived from the eternal chaotic inflationary scenario \cite{Lin86b,Guth00}. The difference is that it is not restricted to a specific epoch but applies
 at any expansion 
epoch.

The next question
 is whether black hole production can have phenomenological consequences. First, recall that inflation is supposed to dilute unwanted components, such as magnetic monopoles, to undetectable
levels. Black holes are no exception. Therefore,  only the post-inflationary epoch could leave a signature of  their production,
so we focus on this.
We 
assume that the universe continues to expand due to the residual value of $\Lambda$ following inflation and
that the topology of each region of spacetime fluctuates according to local dynamics, independent of what happens globally. 
Consequently, we can use the Bousso-Hawking formalism  for
space-time bubbles to calculate the total number of black holes produced from the end of inflation until now.  
For this purpose, we note that the present particle and Hubble horizon radii are  comparable,
\begin{equation}
d_0 \sim   H_0^{-1} \sim  10^{26}\mathrm{m}  \sim 10^{61} \lpl   \, ,
\end{equation}
whereas at the end of inflation, they were
\begin{equation}
d_{\rm end} \sim 1 \, \mathrm{cm}  \sim 10^{33} \lpl  \,, \quad H_{\rm{end}}^{-1} \sim   10^{-26}\mathrm{m} \sim 10^9 \lpl    \,,
\end{equation}
 respectively. This means we can divide the current Hubble horizon into  $(d_0/d_{\rm end})^3 \sim 10^{156}$ bubbles, each 
 with size $H_{\rm{end}}^{-1} \sim 10^9 \lpl$.  However, the number density  has decreased by 
the ratio of the particle horizons cubed
$\sim (d_\mathrm{end}/d_0)^3\sim 10^{-84}$,  so
the number of black holes
produced after inflation is
\begin{equation}
\label{eq:number}
\mathcal{N}_\mathrm{BH} \sim   \mathcal{P} \times  10^{72} .
\end{equation}
For this calculation, we  have set $\ell \sim \lpl$ to obtain $M \sim \mpl$ for the lukewarm case, although it is also possible to consider a more general case with $\ell\neq\lpl$. 
If dark matter consists solely of low-temperature Planckian black holes, we would require approximately 
$\mathcal{N}_\mathrm{BH} \sim 10^{60}$ to account for a mass of about $\sim 10^{52}$ kg within the current Hubble horizon. 
This scenario necessitates a probability $\mathcal{P}\approx \Gamma \equiv \Gamma_{\mathrm{DM}} \sim 10^{-12}$ for each bubble. 
We stress that the rate $\Gamma_\mathrm{DM}$ can only be determined within our framework. For $\Gamma=\Gamma_\mathrm{DM}$,  \eqref{eq:lwrate2} admits a solution for $r_3$ if and only if the causal structure of spacetime corresponds to the line element described in  \eqref{eq:potential}.

We emphasize that the value of $\Gamma_\mathrm{DM}$ is minuscule, implying that the universe produces the full dark matter content over a long period of time while remaining quantum mechanically stable at each epoch.  
 On the other hand,  $\Gamma_\mathrm{DM}$ 
is very large compared to what Mann and Ross \cite{MaR95},  as well as Bousso and Hawking \cite{BoH95,BoH96}, found in their analysis of the production rate of classical black holes after inflation.

\subsection*{Acknowledgements}

P.N. would like to thank the GNFM, the Italian National Group for Mathematical Physics, and the ``Iniziative Specifica FLAG'' of the INFN, the Italian National Institute for Nuclear Physics, Trieste.


\end{document}